\documentclass[conference]{IEEEtran}
\IEEEoverridecommandlockouts
\usepackage{multirow}
\usepackage{cite}
\usepackage{float}
\usepackage{amsmath,amssymb,amsfonts}
\usepackage{algorithmic}
\usepackage{algorithm}
\usepackage{graphicx}
\usepackage{textcomp}
\usepackage{xcolor}
\usepackage{array, diagbox}
\usepackage{graphicx}
\usepackage{subscript}
\usepackage[hyphens]{url}

\def\BibTeX{{\rm B\kern-.05em{\sc i\kern-.025em b}\kern-.08em
    T\kern-.1667em\lower.7ex\hbox{E}\kern-.125emX}}

\begin{document}

\pdfpagewidth=8.5in
\pdfpageheight=11in
\pagenumbering{arabic}

\title{MonteQ: A Monte Carlo Tree Search Based Quantum Circuit Synthesis Framework}
\author{\IEEEauthorblockN{Mulundano Machiya}
\IEEEauthorblockA{
\textit{University of Chicago}\\
Chicago, USA \\
mmachiya@uchicago.edu}
\and
\IEEEauthorblockN{Matt Menickelly}
\IEEEauthorblockA{\textit{Argonne National Laboratory}\\
Lemont, USA \\
mmenickelly@anl.gov}
\and
\IEEEauthorblockN{Paul Hovland}
\IEEEauthorblockA{\textit{Argonne National Laboratory}\\
Lemont, USA \\
hovland@mcs.anl.gov}
\and
\IEEEauthorblockN{Ji Liu}
\IEEEauthorblockA{\textit{Argonne National Laboratory}\\
Lemont, USA \\
ji.liu@anl.gov}
}
\date{October 2025}

\maketitle
\begin{abstract}

Hamiltonian simulation is one of the most promising paths toward quantum advantage. Most prior approaches to Hamiltonian simulation circuit synthesis focus on local rewrite rules and low-level optimizations, and give limited attention to high-level scheduling of Pauli terms under varying constraints. In practice, different simulation algorithms require different orderings of the Pauli terms, yet many prior IR-based methods assume a fixed commutation structure, which limits their flexibility.

We present MonteQ, a novel quantum circuit synthesis framework for Hamiltonian simulation. MonteQ leverages a two-level design that combines low-level synthesis heuristics with an upper-level tree structure to explore sequences of Pauli rotations. To avoid enumerating this factorially large tree, the Monte Carlo Tree Search algorithm serves as workhorse for judiciously exploring promising paths to leaf nodes. 
With this two-level design, MonteQ supports both logical-level and hardware-aware synthesis by selecting different low-level heuristics. It also supports different ordering constraints on the Pauli rotations by adjusting the high-level tree structure. For example, MonteQ can preserve the target unitary by using a directed acyclic graph that records the commutation relations among the Pauli rotations, or it can relax unitary preservation constraint to uncover additional optimization options.
Our experimental results show that MonteQ can achieve an improvement, as measured in CNOT gate counts, of up to 53\% (30\% on average) against state-of-the-art compilers like Rustiq on a set of representative synthesis tasks. 

\end{abstract}
\section{Introduction}
\label{sec:intro}
\noindent Quantum simulation on classical computers is inefficient when studying large problems in many-body physics and quantum chemistry. The difficulty lies in the fact that a quantum state is expressible as a normalized vector in a Hilbert space, the dimension of which grows exponentially with the size of the quantum system; thus, representations of a quantum state generally require an exponentially large vector of complex numbers. 
However, quantum circuits capable of simulating some quantum system can be defined in the same Hilbert space and can serve as a more compact representation.
The number of qubits (quantum bits) in a circuit necessary for such simulation tasks typically grows linearly in the size of the simulated quantum system.
For this reason, quantum computers have been presented as universal simulators for quantum systems~\cite{lloyd1996universal}.  
However, noisy intermediate-scale quantum (NISQ)-era quantum computers suffer from an error rate that often increases proportionally to circuit size. 
Minimizing the number of two-qubit gates appearing in a circuit is particularly important in this regard because two-qubit gates tend to have a higher error rate than their single-qubit counterparts~\cite{quantum7010002}, necessitating quantum circuit optimization.\par
\noindent Pre-exisiting compilers for Hamiltonian simulation leverage different optimization techniques to reduce two-qubit gate counts. 
Synthesis directly into Pauli networks like in QuCLEAR\cite{10946817} and Rustiq~\cite{ de2024rustiq}, and Pauli string reordering for improved local gate cancellation ~\cite{10.1145/3503222.3507715, Jin:2023mil} are two well-researched methods that aim for reduction of two-qubit gates.
Greedy search strategies are typically employed in previous Pauli network synthesis works. For example, Rustiq opts to first sort Pauli strings by their Pauli weights, defined as the number of non-identity terms, and QuCLEAR can only carry out its recursive CNOT tree construction as far as the number of qubits.  
It has not been sufficiently explored if such greedy search strategies in these contexts restrict opportunities for global optimization. 
MonteQ addresses this by recognizing the existence of a set of (intractably large) equivalent Pauli networks, and seeks to sample these networks intelligently, selecting one from the sample that exhibits the lowest count of CNOT gates.
MonteQ expresses the problem of searching equivalent circuits as one of exploring paths through a finite search tree, which is explored via a method built around Monte Carlo Tree Search (MCTS)~\cite{doi:https://doi.org/10.1002/9781119815068.ch19}. 

\noindent Another challenge is that different optimization targets often require different synthesis algorithms. For example, hardware-aware synthesis incorporates device connectivity into the algorithm design. For Hamiltonians in many-body physics simulation, the system topology can also be included in the algorithm design. Redesigning the problem for each optimization target and each class of Hamiltonian is difficult. Therefore, our proposed MonteQ framework uses a two-level design: the low-level synthesis heuristics tailored to different optimization objectives, while the upper-level tree structure uses Monte Carlo Tree Search to explore different ordering strategies of the Pauli strings. The MonteQ framework supports multiple optimization targets and allows new designs for additional targets to be added easily.\par
\noindent In this paper, we demonstrate MonteQ's circuit synthesis framework with a focus on quantum simulation. The framework is characterized by four primary features:
\begin{itemize}
    \item We introduce a heuristic that induces a tree-like structure over the space of equivalent circuits, and formalize the synthesis problem as a Markov Decision Process.
    \item Monte Carlo Tree Search (MCTS) is the core search method in MonteQ and is used to explore spanning trees of the heuristic-defined tree, enabling efficient search over a large space of feasible quantum circuits.
    \item The framework is sufficiently flexible to support the use of alternative heuristics for tree generation besides the ones we experiment with in this paper, as well as different high-level Pauli string ordering strategies that either preserve the circuit’s underlying unitary or modify the unitary according to an optimization objective.
    \item The MonteQ framework provides a flexible tradeoff between solution quality and runtime by adjusting the number of MCTS iterations, allowing the balance to be tuned based on user needs.
\end{itemize}
The remainder of the paper is organized as follows. 
Section~\ref{sec:background} details the background necessary to describe the circuit synthesis problem as an optimization problem and the related works. Section~\ref{sec:problem formulation} formulates the circuit synthesis problem underlying MonteQ.
Section~\ref{sec:design} discusses the overall design of MonteQ, a method designed to tackle this optimization problem. 
Section~\ref{sec:evaluation} presents preliminary results concerning the performance of MonteQ on a set of relevant benchmarks. 
Finally, we provide concluding remarks in Section~\ref{sec:conclusion}. 

\section{Background and Related works}
\label{sec:background}
\subsection{Quantum Simulation}
\label{subsec:sim}
\noindent Quantum simulation involves the use of a quantum system to study the properties of another separate quantum system.  
Many areas of interest within quantum chemistry and physics are being studied through the lens of quantum simulation.
The application of variational methods like qubit coupled cluster (QCC)~\cite{doi:10.1021/acs.jctc.8b00932} and unitary coupled cluster (UCC) theory~\cite{romero2018strategies, PhysRevA.98.022322}, and unitary time evolution and observable measurement ~\cite{Fauseweh2024-ni} are all clear examples. 
In most of these studies, the main goal is the implementation of an operator $U$ to a reference state. Using a mapping to qubits~\cite{Chiew:2021gtk} followed by an approximation scheme such as standard trotterization~\cite{trotter1959trotter} or a randomized trotterization such as that employed in QDrift~\cite{PhysRevLett.123.070503}, the operator can be approximated as a sequence of Pauli rotations
\begin{equation}U \approx \Pi_j e^{-i \theta_j P_j},\label{eq:U}\end{equation}
where
$P_j = \bigotimes^{n}_{i=1}\sigma_i$ is a n-qubit tensor product of Pauli operators, $\sigma_i \in \{I, X, Y, Z\}$;  we refer to $P_j$ as a \emph{Pauli string}. 
The terms $e^{-i \theta_j P_j}$ are the associated \emph{Pauli rotations}.\par
\noindent To implement $U$, a circuit needs to be synthesized in a universal gate set that can be executed on a quantum computer. 
In this paper, we employ Clifford and Rz as a universal gate set. Our primary goal is to minimize the number of CNOT gates appearing in the constructed Clifford subcircuits. 

\subsection{Clifford Circuits}
\label{subsec:Cliff}
\noindent Clifford circuits are quantum circuits composed of CNOT $CX_{ij}$ ($i$ indexes the control qubit and $j$ indexes the target qubit), Hadamard $H$, and Phase $S$ gates. 
Clifford circuits have the property that they normalize the Pauli group~\cite{Grier2022classificationof}. That is, given an n-qubit Pauli string $P_1$ and an n-qubit Clifford circuit $CL^n$, 
\begin{equation}P_2 = CL^{n}P_1CL^{n\dagger}.\label{eq:normalizer}\end{equation}
This property can be extended to the exponentiation of those Paulis, that is 
\begin{equation}e^{-i\theta P_2} = CL^{n}e^{-i\theta P_1}CL^{n\dagger}.\label{eq:rotation_normalizer}\end{equation} 
In words, any Pauli rotation can be mapped to a single-qubit rotation conjugated by a Clifford circuit.\par
\begin{figure}[htbp]
\centerline{\includegraphics[width=\linewidth]{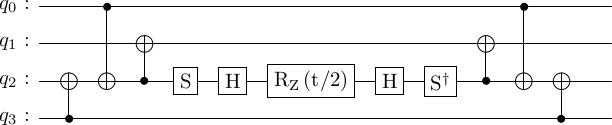}}
\caption{Clifford+Rz Representation of $e^{-iZYXZ(t/4)}$. $R_z(t/2) = e^{-iZ(t/4)} $}
\label{fig:example} \end{figure} 
\noindent For example, Fig~\ref{fig:example} illustrates $e^{-iZ_3Y_2X_1Z_0\frac{t}{4}} = CL^{4}e^{-iI_3Z_2I_1I_0 \frac{t}{4}}CL^{4\dagger}$. 
The normalization property \eqref{eq:normalizer} implies that for any Pauli string there exists a Clifford circuit that maps it to a second Pauli string with Pauli weight 1. 
Furthermore, the relationship \eqref{eq:normalizer} can be treated as weakly commuting for each Pauli string or rotation, that is,  $P_2 = CL^{n}P_1CL^{n\dagger} \rightarrow CL^{n\dagger}P_2 = CL^{n\dagger}CL^{n}P_1CL^{n\dagger}  \rightarrow CL^{n\dagger}P_2 = P_1CL^{n\dagger} $~\cite{de2024rustiq}.\par 

\noindent Table~\ref{tab: conj} details the result of conjugating individual Pauli operators with $S$ and $H$:~\cite{reid2024simplemethodcompilingquantum}.

\begin{table}[ht] 
\caption{Conjugations of Pauli operators with $H$ and $S$}
    \begin{center}
        \begin{tabular}{ |c||*{4}{c|}|}
            \hline
             & $\pm I$ & $\pm X$ & $\pm Y$ & $\pm Z$ \\
            \hline
            $ H$ & $\pm I$& $\pm Z$ & $\mp Y$ & $\pm X$\\
            \hline
            $ S$ & $\pm I$ & $\pm Y$ & $\mp X$ & $\pm Z$\\
            \hline
            
            \hline
        \end{tabular}
    \end{center}

 \label{tab: conj}
 \end{table}

\noindent Table~\ref{tab:conjugate} shows the result of conjugating Pauli strings of the form $P = \pm \sigma_i \otimes \sigma_j$ ($\sigma_i, \sigma_j\in \{I,X,Y,Z\}$) with $CX_{ij}$~\cite{10946817}.
In the remainder of this paper, we omit the tensor symbol in such pairs of Pauli operators for brevity whenever the context is clear.
\begin{table}[ht] 
\caption{Conjugation Identities for $CX_{ij}$}
    \begin{center}
        \begin{tabular}{ |c||*{4}{c|}|}
            \hline
            \diagbox[]{$P_i$}{$P_j$} & $I$ & $X$ & $Y$ & $Z$ \\
            \hline
            $ I$ & $I_iI_j$ & $I_iX_j$ & $Z_iY_j$ & $Z_iZ_j$\\
            \hline
            $ X$ & $X_iX_j$ & $X_iI_j$ & $Y_iZ_j$ & $\mp YY$\\
            \hline
            $ Y$ & $Y_iX_j$ & $Y_iI_j$ & $\mp X_iZ_j$ & $X_iY_j$\\
            \hline
            $ Z$ & $Z_iI_j$ & $Z_iX_j$ & $I_iY_j$ & $I_iZ_j$\\
            \hline
        \end{tabular}
    \end{center}

 \label{tab:conjugate}
\end{table}

 \begin{table*}[h!] 
\caption{Comparison of related Hamiltonian simulation synthesis methods}
    \begin{center}
        \begin{tabular}{|c|c|c|c|c|}
            \hline
            Name & Logical Synthesis & Hardware-aware Synthesis & Unitary Preserving & Unitary Modifying \\
            \hline
            TKet~\cite{Sivarajah:2020lfo}& \checkmark & \checkmark & \checkmark &\\
            \hline
            Tetris~\cite{Jin:2023mil}& & \checkmark && \checkmark\\
            \hline
            Rustiq~\cite{de2024rustiq}& \checkmark & & \checkmark & \checkmark\\
            \hline
            QuCLEAR~\cite{10946817}& \checkmark & & \checkmark & \\
            \hline
            MonteQ (this work) & \checkmark & \checkmark& \checkmark & \checkmark \\
            \hline
            
        \end{tabular}
    \end{center}
\vspace{-5mm}
 \label{tab:related_works}
\end{table*}

\subsection{Representation of Pauli Strings} 
\label{subsec:string_rep}
\noindent To construct a circuit that implements an operator in the form of~\eqref{eq:U}, there must first be a way to represent the sequence of Pauli rotations. It is clear that a conjugation of Clifford circuits around Pauli rotations,$e^{-i\theta P_2} = CL_{n}e^{-i\theta P_1}CL_{n}^{\dagger}$, is equivalent to the Pauli rotation of the original Pauli string conjugated by the Clifford circuit, $e^{-i\theta P_2} = e^{-i\theta CL_{n}P_1CL_{n}^{\dagger}}$. Thus, given a sequence of Pauli rotations, it suffices to store only the Pauli strings and values of the corresponding $\theta$ parameters without loss of information. 
This observation enables the vectorization of the Pauli strings using a 2-bit encoding~\cite{de2024rustiq, PhysRevA.70.052328}:
\[X = [1 | 0], \: Y=[1 | 1],\: Z=[0 |1],\: I=[0|0],\: + = 0,\: - = 1\]
We refer to the left side of the encoding as the "$x$"-side and we refer to the right side of the encoding as the "$z$"-side. 
Using these encodings we can represent any Pauli string as a vector. 
For example,
\[XZZX = [\,1\,0\,0\,1\,|\,0\,1\,1\,0\,|\,0]\]
The leftmost portion of the vector contains the $x$-sides for each Pauli in the string, the middle portion contains the $z$-sides of each Pauli and the rightmost portion contains the sign. 
This means an $n$-qubit Pauli string may be represented by a vector $v$ of length $2n+1$. Furthermore, there exists a vectorized method for performing every conjugation identity in Tables ~\ref{tab: conj} and ~\ref{tab:conjugate}.
In particular, 
given a vector $v$ representing an $n$-qubit Pauli string $P$, the application of a Clifford, $CL_{n}PCL_{n}^{\dagger}$, employs one of the following rules depending on the gate type~\cite{de2024rustiq, PhysRevA.70.052328}:\\
\\
\textbf{Hadamard $H_i$ on qubit $i$ } :  $v[i] = v[n + i]$;\quad $v[n + i] = v[i]$;\quad $v[2n] = v[2n] \oplus (v[i]*v[n+i])$; \quad $v[k] = v[k], k\neq \{i, n+i, 2n\}$;\quad \\
\\
\textbf{Phase $S_i$ on qubit $i$} : $v[n + i] = v[i] \oplus v[n+i]$;\quad $v[2n] = v[2n] \oplus (v[i]*v[n+i])$;\quad $v[k] = v[k], k\neq \{n+i, 2n\}$;\quad\\
\\
\textbf{CNOT $CX_{ij}$ on control $i$ and target $j$} : $v[j] =v[i]\oplus v[j]$;\quad $v[n+i] = v[n+i] \oplus v[n+j] $;\quad $v[2n] = v[2n]\oplus (v[i]*v[n+i])*(v[j]\oplus v[n+i]\oplus 1)$;\quad$v[k] = v[k], k\neq \{j,n+i, 2n\}$;\quad \\
\\
A sequence of $K$ Pauli strings can be represented as a $K \times (2n + 1)$ matrix, which we refer to as a length $K$ \emph{Pauli word}. For example,

 \[\begin{bmatrix}
  XZZX \\
  YXXY
  \end{bmatrix}
  =\begin{bmatrix}
  1& 0 & 0 & 1 | &0&1&1&0 |&0\\
  1&1&1&1 | & 1&0&0&1 |&0 
  \end{bmatrix}
  \]
\\
 The stated conjugation rules for vectors $v$ can be applied naturally to each column of a Pauli word.

\subsection{Related Works}
\noindent Efficient compilation and synthesis of Hamiltonian simulation circuit is an active research area. Table~\ref{tab:related_works} summarizes the synthesis methods compared in our experiments. As shown in the table, existing approaches target different objectives, such as hardware awareness and unitary preservation, and most prior works do not address all of these objectives within a single framework. Due to its two-level design, MonteQ can naturally support multiple optimization targets.

\noindent Local gate cancellation has been widely studied in prior work, since gates in Pauli-string subcircuits can cancel between similar Pauli strings. Li et al. proposes Paulihedral~\cite{10.1145/3503222.3507715} which leverages Pauli-based intermediate representation (IR) to enable gate cancellation and hardware-aware synthesis. Paulihedral also introduces a scheduling algorithm that orders Pauli strings to maximize cancellation opportunities. Jin et al. propose Tetris~\cite{Jin:2023mil}, which introduces a refined IR that more explicitly accounts for gate cancellation, hardware-aware synthesis, and SWAP insertion on devices with limited connectivity. Tetris presents a similarity metric and a scheduling algorithm for the order of the Pauli strings. Both Paulihedral and Tetris rely on reordering blocks of Pauli strings that are not necessarily commuting, enabling optimizations but not preserving the unitary.

\noindent Beyond approaches based on local gate cancellation, Pauli network synthesis methods have received increased attention because they can produce more efficient circuits. Rustiq~\cite{de2024rustiq} proposes a bottom-up synthesis heuristic that evaluates the cost of two-qubit Clifford blocks. It uses a greedy cost function to compare different block choices, and the full Pauli network is constructed by composing these blocks.  PHEONIX~\cite{yang2025phoenix} employs a heuristic for finding two qubit Clifford operators in sequence to maximally simplify the circuit. QuCLEAR~\cite{10946817} propose a heuristic CNOT tree synthesis method together with a Clifford absorption method to resolve the final Clifford subcircuit.

\noindent In addition to these studies, TKet~\cite{Sivarajah:2020lfo} compiler resynthesize quantum circuits into ZX diagram representation. They leverage the local commutation rules of the Paulis to resynthesize the Pauli gadgets~\cite{cowtan2019phase_phasegadget}. Glos et al.~\cite{glos2024generic} propose a Clifford Executive Representation for synthesizing circuits on hardware with limited connectivity. Kernpiler~\cite{decker2025kernpiler} proposes partial Trotterization and incorporates the Trotter error analysis in the synthesis process. In our experiments, we compare with TKet, QuCLEAR, and Rustiq implemented in Qiskit in both unitary preserving and unitary modifying settings. We also include comparisons with TKet, Tetris and Qiskit hardware transpiler in the hardware-aware experiments.


\section{Circuit Synthesis Problem Formulation}
\label{sec:problem formulation}
\noindent In this section, we formulate the circuit synthesis problem underlying MonteQ. In particular, we model the synthesis problem as a sequential optimization problem that can be framed as a Markov Decision Process (MDP)~\cite{doi:https://doi.org/10.1002/9781119815068.ch2}.
\subsection{Circuit Synthesis as a Sequential Optimization Problem}
\label{circ op}
\noindent We now discuss how to model the problem of circuit synthesis as one amenable to sequential optimization.
We intend to synthesize a circuit with specified unitary 
$$U= e^{-i\theta_1P_1}e^{-i\theta_2P_2}...e^{-i\theta_kP_K}.$$ 
By our prior observations concerning Clifford circuits (in \eqref{eq:rotation_normalizer}),
we can express
\begin{equation}e^{-i\theta_1P_1}= CL_1^n e^{-i \theta_1 P_1^{'}} CL_1^{n\dagger}\label{eq:first_pauli}
\end{equation}
for the first Pauli string $P_1$ appearing in $U$. 
In turn, this new rotation $e^{-i \theta_1 P_1^{'}}$ can be expressed entirely in terms of Clifford and Rz gates, 
$$e^{-i \theta_1 P_1^{'}}= CL_1^{n'}R_z\left(\theta_1\right)CL_1^{n'\dagger}.$$
Then, the tail Clifford $CL_1^{n'\dagger}$ can be commuted to the end of the sequence of Pauli strings, yielding
\begin{equation}\label{eq:intermediate_U}
U = CL_1^{n}R_z\left(\theta_1\right)e^{-i\theta_2P'_2}...e^{-i\theta_kP'_K}CL_1^{n\dagger}.
\end{equation}
Observe that this commutation necessarily updated the Pauli strings $P_2,\dots,P_K$ to new Pauli strings $P_2^{'},\dots,P_K^{'}$, and we updated the notation so that the Clifford gate associated with what was originally $P_1$ is again simply $CL_1^n$, although this is generally distinct from the Clifford gate appearing in \eqref{eq:first_pauli}. 
  

\noindent This procedure is repeated an additional $K-1$ times to produce a full Clifford+Rz circuit that implements $U$, that is, 

\begin{equation}\label{eq:synthesized_U}
U = CL_1^nR_z\left(\theta_1\right)...CL_K^nR_z\left(\theta_K\right)CL_K^{n\dagger}...CL_2^{n\dagger}CL_1^{n\dagger}.
\end{equation}


\noindent Thus, after this process $U=U'U_{CL}$, where $U'$ is the leading Clifford+Rz circuit and $U_{CL}$ is the trailing Cliffords after the last Rz. The number of ways to synthesize $U$ by following this procedure is factorial in $K$. 
In particular, we can select the order in which we reduce a Pauli rotation to an equivalent set of Clifford and Rz gates and subsequently commute the tail Clifford to the end of the circuit; we refer to this step as \emph{implementation of the Pauli rotation}. 
Critically, however, because the commutation modifies the remaining Pauli rotations at the end of each of the $K$ implementations (see \eqref{eq:intermediate_U}), the Pauli weights of each string left remaining in the Pauli word change between each step of the procedure. 
Thus, a myopic selection rule that always greedily decides to implement next the Pauli string with the least Pauli weight may very well result in Clifford circuits with unnecessarily large CNOT counts. 


\subsection{Markov Decision Process}
\noindent This aversion to myopic decision making motivates a desire to employ methods for sequential optimization. Towards this end, we formally describe circuit synthesis as a Markov decision process (MDP) by identifying the state space, the action space, the transition function, and the reward. 
\begin{itemize}
    \item[]\textbf{State} - The states, each denoted $s$, correspond to both a Pauli word and the sequence of implementations that led from an initial state to the given Pauli word. The initial state is always the initial Pauli word representation of $U$ where none of the Pauli strings have been implemented. 
    The terminal states correspond to Pauli words of length 0, that is, the set of circuits already expressed entirely in terms of Clifford and Rz gates. 
    \item[] \textbf{Action} - At each stage, the action $a$ is defined as the index of the Pauli string we choose to implement next.  
    Observe that the maximum number of available actions in each stage of the MDP is equal to the length of the Pauli word.
    \item[] \textbf{Transition function} - The transition function $\Phi(s,a)$ is simply the abstract function that returns the resulting Pauli word of length $K-1$ upon taking the action of implementing a Pauli string from a Pauli word of length $K$. Observe that because this transition function is deterministic by definition, the MDP is deterministic.  
    \item[] \textbf{Reward} - The reward $r(s,a)$ for taking an action $a$ at a state $s$ is the negative of the number of CNOTs required to implement the Clifford circuit for the chosen Pauli string. Hence, the cumulative reward upon reaching the terminal state is the sum of the CNOTs appearing in all $K$ tail Cliffords. 
\end{itemize}

\section{MonteQ Design}\label{sec:design}
\begin{figure*}[htbp]
\centerline{\includegraphics[width=\linewidth]{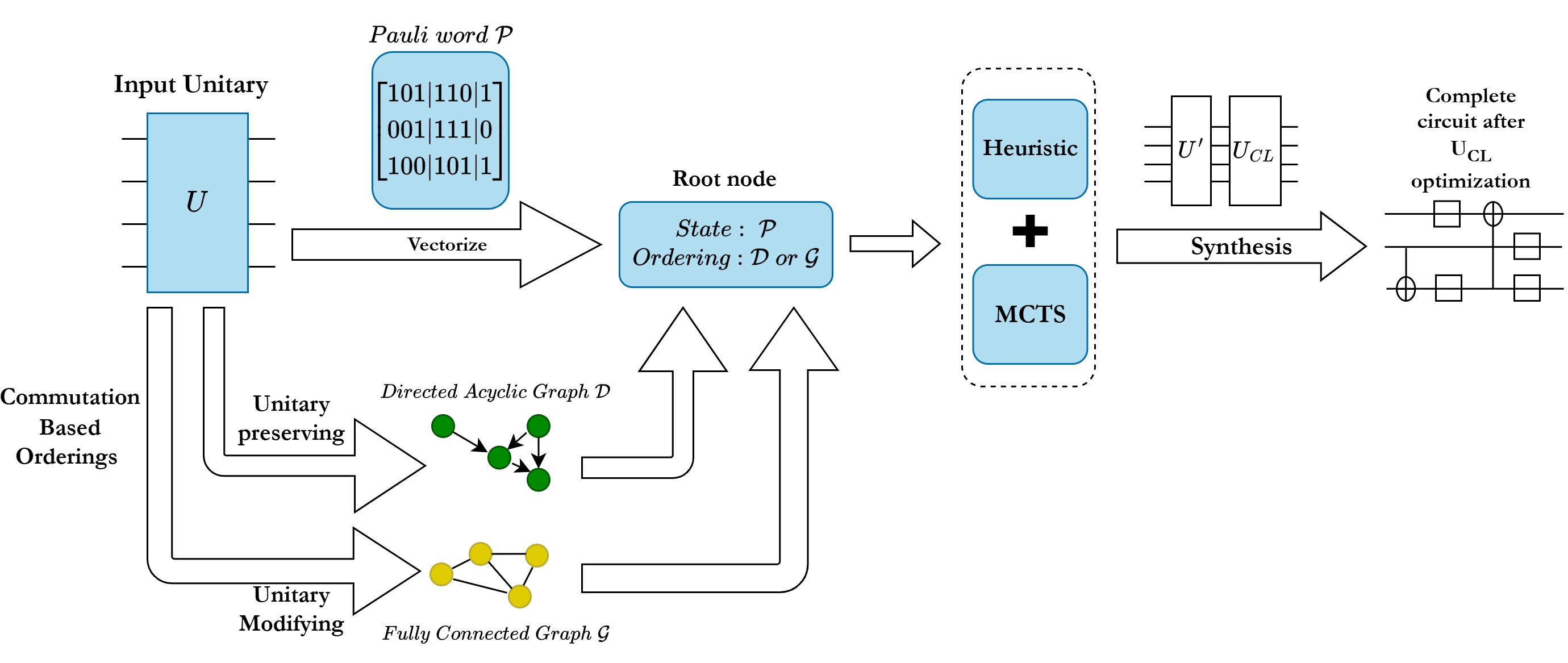}}
\caption{Overview of MonteQ Framework}
\label{fig: Frame} \end{figure*}
\subsection{Overview}
\noindent To summarize, the problem MonteQ aims to address is: given a unitary $U$, what is an equivalent circuit of least (or, at least, low) CNOT gate count?
We note that how an implementation, that is any action of the MDP, is performed still involves degrees of freedom. 
Thus, MonteQ requires a heuristic for performing an implementation as an input. 
Based on prior discussion, the implementation heuristic applied to a Pauli word of length $K$ must return a Pauli word length $K-m$ for $m \geq 1$, as well as the $n$-qubit Clifford+Rz circuit.\par 
\noindent The choice of implementation heuristic that one provides MonteQ can be motivated by, for instance, a desire for unitary preservation. 
Indeed, the set of available actions out of any state within the MDP framework can be determined in one of two ways: 
\begin{enumerate}
    \item \textbf{Unitary Preserving} - the sequence of Pauli strings is represented as a directed acyclic graph (DAG). 
    The original indices of the Pauli strings define the nodes of the DAG. If two Pauli strings anticommute, a directed edge extends from the node of the Pauli string that comes first in the sequence to the node of the second Pauli string. 
    The nodes with in-degree zero commute and can be implemented in any order. The set of all nodes with in-degree zero is called a \emph{front layer} ~\cite{de2024rustiq}. Synthesizing via this method preserves the unitary while also exploiting possible commutative reordering. 
    This front layer of nodes can be used to define the set of available actions at each stage of our MDP.  
    As Pauli strings are implemented, the corresponding nodes and adjacent edges are removed, opening up new front layers as the tree is built.
    \item \textbf{Unitary Modifying} - In an unitary modifying setting, the same DAG is implicitly a complete directed graph, that is, the Pauli word is treated as if every Pauli string commutes. This allows any arrangement of the Pauli strings and so our original unitary is not preserved.
\end{enumerate}
\noindent In what follows, we propose to use Monte Carlo Tree search to effectively identify spanning trees of this DAG, iteratively constructing a search tree capable of both representing equivalent circuits and tracking objective progress. 

\subsection{Monte Carlo Tree Search}
\begin{figure}[htbp]
\centerline{\includegraphics[width=\linewidth]{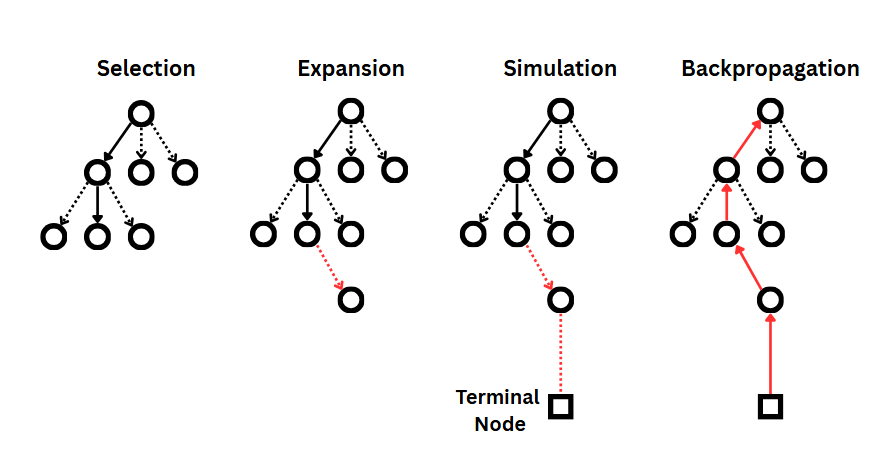}}
\caption{Illustration of Monte Carlo Tree Search}
\label{fig:mcts} \end{figure}
\noindent Monte Carlo Tree Search (MCTS) is a framework for solving sequential decision problems, such as the problem of circuit synthesis. 
More specifically, MCTS methods typically apply to discrete sequential decision problems where out of each discrete state, a finite number of actions are available. 
In such a setting, we can associate all states with nodes of a graph and all actions with directed edges between states. 
Then, given an initial state corresponding to a root node, the graph union of all paths through an MDP to terminal states (leaf nodes) defines a tree.
By maintaining an estimate of a value for any given node of the tree, an MCTS method judiciously and iteratively explores this (often very large) search space by initializing a root node corresponding to the initial state, and then growing a tree starting from this root node by adding one node adjacent to the tree per iteration.
At every iteration of an MCTS method, each node $s$ in the current tree maintains a dynamic estimate of the value of the node, $V(s)$.
In summary, the 4 steps per iteration of MCTS can be described as follows, and are illustrated in Fig~\ref{fig:mcts}: 
\begin{enumerate}
    \item \textbf{Selection} - Beginning from the root node, some policy is employed to determine an edge (action) to traverse to a new node (state). 
    This policy generally 1) is based on the current estimates of the value of each adjacent node, and 2) aims to balance exploration of unvisited paths through the MDP with exploitation of paths with seemingly high values. 
    This selection step is repeated until a leaf node of the current tree is reached. The selection policy we employ is based on the Upper Confidence Bound for Trees (UCT), that is, at state $s$, we choose
\begin{equation}a^* \gets \displaystyle\arg\max_{a\in A} r(s,a) + V(\Phi(s,a)) + \mu\sqrt{\frac{\log N_s}{N_{s,a}}}\label{eq: UCT}\end{equation}

\noindent The former term of \eqref{eq: UCT} aims to maximize the value of the terminal node that the tree search will probe, while the latter term encourages non-myopic exploration; $N_s$ denotes the number of simulations (defined in Step 3) run out of state $s$, $\mu$ is the exploration factor, which is set to $\sqrt{2}$ for this work, and $N_{s,a}$ denotes the number of times action $a$ was previously chosen out of node $s$. \par

    \item \textbf{Expansion} - Assuming the leaf node from the selection step does not correspond to a terminal state, the leaf node is \emph{expanded}; that is, the policy from the selection step is again applied to determine an action (edge) and new state (adjacent node) which are added to the tree. 
    \item \textbf{Simulation} - From the newly expanded node, a \emph{rollout} policy, that is, a (generally heuristic) policy that is easy to compute and does not rely on value estimates, is run to simulate a path to a terminal state of the MDP. 
    In our implementation, we employ a purely greedy rollout policy that simply chooses, at each state, to implement the remaining Pauli string of least Pauli weight. 
    \item \textbf{Backpropagation} - Starting from the terminal state of the simulated path, an estimate of the value $V(s)$ of every state $s$ along the path is computed and backpropagated to the root of the tree. 
    In our implementation, we simply let $V(s)$ be a running average of the reward at any terminal node reachable from $s$.
    We recall from the selection step that this backpropagation step simultaneously increments $N_{s,a}$ as appropriate. 
\end{enumerate}
\noindent In problem settings where the size of the full tree representing the finite discrete MDP is intractably large to fully enumerate, it is standard practice to let MCTS iterate over these four steps for some fixed amount of time or some fixed number of iterations. 
As we shall see in Section~\ref{sec:design}, the MDP model we employ in this work is intractably large for reasonably sized Pauli words, and so we will develop an MCTS algorithm with an assumed maximum number of iterations.

\subsection{Heuristic Design}
\noindent A strength of MonteQ is the flexibility of its implementation heuristic. 
A noteworthy concern is that the performance of MonteQ may depend heavily on the availbility of sophisticated heuristics. 
We assuage this concern by presenting two fairly simple heuristics and demonstrating their performance within MonteQ in the subsequent section.\\
\\
\noindent\textbf{Logical Regime Greedy Heuristic}:
\noindent The \emph{logical regime} assumes all-to-all connectivity of our qubits. This regime is useful for studying logical-level circuit optimization and for targeting hardware platforms with native all-to-all connectivity, such as trapped-ion systems~\cite{Chen2024benchmarkingtrapped}. Because of the lack of connectivity constraints, more paths for reduction open up as CNOTs can be placed between any two qubits. We design a \emph{greedy implementation heuristic}, algorithm \ref{algo 1}, based on the heuristic used in Rustiq to construct a Pauli network~\cite{de2024rustiq}.\par
\label{heuristic}

\begin{figure}[htbp]
\centerline{\includegraphics[width=25mm]{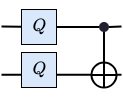}}
\caption{Clifford blocks used to construct a Pauli network. Here, $Q\in \{I, H, S\}.$}
\label{fig:block} \end{figure}

\noindent We construct a Pauli network using \emph{Clifford blocks}, illustrated in  Fig~\ref{fig:block}. Observe that the only way to reduce the Pauli weight of a given Pauli string is to conjugate a pair of Pauli rotations with a CNOT gate. 
Table~\ref{tab:conjugate} shows that conjugation with $CX_{ij}$ applied to any of the pairs $X_iX_j,\, Z_iZ_j,\, Y_iX_j,\, Z_iY_j$ will lead to a reduction in Pauli weight. 
We refer to these four pairs as \emph{reducible pairs (RPs)}. 
Any pair of Pauli rotations can be transformed into an RP by conjugating the pair with another Clifford. 
Achieving this transformation into an RP is exactly the purpose of the abstract $Q$ in our Clifford blocks.
We note that conjugating any of the pairs $X_iI_j,\, I_iZ_j,\, Y_iI_j,\, I_iY_j$ with $CX_{ij}$ will result in an undesirable increase in Pauli weight.
We refer to these four pairs as \emph{increasing pairs (IPs)}. \par
\noindent Our greedy implementation heuristic examines a given Pauli string and cycles through all the qubit pairs in the Pauli string.
The heuristic intentionally ignores pairs that contain $I$ on one of the qubits, because this pair cannot be reduced by a Clifford block. 
For each qubit pair $i,j$, the heuristic generates a list of $Q$ pairs in a Clifford block that transform the Pauli rotations in the pair into an RP. 
To choose the most appropriate $Q$ pair for $i,j$, we define a metric, which we call the \emph{benefit},

\begin{equation}
    benefit_{ij} = \#RP_{ij} - \#IP_{ij}
\end{equation}

\noindent That is, the benefit is defined as the additional number of RPs that are generated across the entire Pauli word by conjugation with a Clifford block defined by a given $Q$ pair, reduced by the number of IPs that are generated across the entire Pauli word by applying the same Clifford block. 
The $Q$ pair with the highest benefit is selected for the qubit pair $i,j$.\par
\noindent Greedily, the heuristic selects the most beneficial $Q$ pair for  every $i,j$ qubit pair in the Pauli string, provided the qubit pair does not involve an identity.
If more than one $Q$ pair has the same benefit, the heuristic selects the pair that leads to the largest number of RPs, breaking ties arbitrarily.
Once the most beneficial $Q$ pair is identified, they are combined with a CNOT to create a Clifford block. This Clifford block is conjugated against the Pauli word using the conjugation rules provided in Subsection ~\ref{subsec:Cliff}.  
This process is repeated until the original Pauli string is implemented. 
The heuristic returns the reduced Pauli string, and a list encoding the Clifford blocks and single-qubit Pauli operations required to implement the Pauli string.
As we have noted previously, the correct combination of $H$ and $S$ can transform any Pauli rotation to a Pauli $Z$. Thus, transforming single-qubit rotations into Pauli $Z$ does not add to the CNOT count. This heuristic is presented pseudocode in Algorithm~\ref{algo 1}. Within Algorithm~\ref{algo 1}, \emph{prune} is an abstract method that extracts the implemented Pauli strings.\\

\begin{algorithm}
\label{algo 1}
  \caption{Logical Regime Greedy Heuristic} 
  \begin{algorithmic}[1]
  \label{algo 1}
  \REQUIRE length k Pauli word $\Omega_k$, specific Pauli string index $ndx$, allowed Pauli word indexes ndxs
  \STATE $ndxs \leftarrow \Omega_k[ndx].indexes() $
  \STATE $reduction = []$
  \WHILE{$Pauli\_weight(\Omega_k[ndx])\:!= 1$ }
    \STATE $cliffs = []$
    \FOR{all $\langle i,j\rangle$ in $ndxs$}
        \STATE $ctrl = \Omega_k[ndx][i],\;targ =\Omega_k[ndx][j]$
        \IF{$ ctrl == I$ or $ targ == I$}
            \STATE skip
        \ENDIF
        \STATE $pair\_cliffs \leftarrow gen\_cliffs(ctrl, targ)$
        \STATE $cliffs.append(most\_benefit(pair\_cliffs,\Omega_k))$
    \ENDFOR
    \STATE $cliff\_pair \leftarrow most\_benefit(cliffs, \Omega_k))$
    \STATE $block \leftarrow cliff\_pair + CX_{ij}$
    \STATE $\Omega_k \leftarrow block\;\Omega_k\;block^{\dagger}$
    \STATE $reduction.append(block)$
  \ENDWHILE
  \STATE $\Omega_{k-m}, single\_q \leftarrow prune(\Omega_k, ndxs)$
  \STATE $reduction.append(single\_q)$
  \RETURN $\Omega_{k-m}, reduction$
  \end{algorithmic}
\end{algorithm}


\noindent \textbf{Hardware-Aware Heuristic}
The greedy implementation heuristic for the logical regime is not unique. As long as a heuristic can be designed that takes in a Pauli word of length $K$ and returns a reduced Pauli word and the $n$-qubit circuit that implements the removed strings, the heuristic is valid. Thus, we can introduce a simple \emph{hardware-aware heuristic}, algorithm \ref{algo 2}, inspired by the algorithms presented in ~\cite{10.1145/3297858.3304023, Zou:2024sij}.\par
\noindent A Pauli word of length $K$ can be divided into sections $\Omega_K = [X|Z|S]$ where $X$ and $Z$ are $K \times n$ matrices representing all the x-sides and z-sides, respectively, of each vectorized Pauli string. $S$ is a $K\times 1$ matrix representing the sign of each string. An \emph{occupancy matrix}, $O$, can be formed by taking a logical OR of the entries in $X$ and $Z$; that is, $O = X  || Z$.
The occupancy matrix indicates which qubits are occupied by non-identity Paulis. Given an occupancy matrix, a connectivity graph $C$ specifying the connectivity pattern of qubits, corresponding edge weights $D$ describing the distance between connected qubits, a length $K$ Pauli word, and an index for a Pauli string to implement, our hardware-aware heuristic returns a length $K-m$ Pauli word ($m\geq 1)$ along with the Clifford+Rz gates that implemented the Pauli string.\par
\noindent The primary difference between our hardware-aware heuristic and the logical regime heuristic lies in the initialization. 
In the logical regime, every qubit is connected to every other qubit, that is, $C$ is a complete graph. 
However, some hardware configurations have unconnected qubits. Therefore, it is wisest to eliminate -- that is, replace with an $I$ -- all Pauli operations lying on qubits that exhibit the largest distance (as per $D$) from other occupied qubits. 
This is meant to avoid removing existing connections, as well as to avoid increasing the number of SWAPs, $SWAP = CX_{ij}CX_{ji}CX_{ij}$. \par
\noindent Our hardware-aware heuristic identifies the non-identity qubits with the largest distance from any other non-identity qubit in $C$. The goal is to eliminate the Paulis acting on those qubits with the largest distance. As discussed, reducible pairs are the only combinations that can produce an identity by conjugation. So, again, creating reducible pairs is necessary. However, unlike the logical regime, we need Paulis on specific qubits to be converted to I. To do this the heuristic generates a list of all the Clifford operations - combinations of a CNOT, and one or more S and H gates - that will eliminate the Paulis on those specific qubits upon conjugation.  
We define a distance metric
\begin{equation}
    dist = \displaystyle\displaystyle\sum_n[(O\times D) .*O]_{kn},
\end{equation}

\noindent and select a Clifford operation from the list that most reduces $dist$. In words, $dist$ is the sum of the matrix elements of $[(O\times D) .*O]$ for the $k$-th Pauli string, which  we are attempting to implement. Here, $\times$ represents matrix multiplication and $.*$ represents element wise multiplication. In the pursuit of minimizing the distance metric, it is possible to have a set of operations that improve the metric but do not reduce the Pauli weight of the selected Pauli string. 
Thus, the hardware-aware implementation heuristic also keeps track of prior states of the Pauli word. 
If a state is repeated, this implies that the heuristic is caught in a loop because it chose a move that reduces the distance metric, but does not reduce the Pauli weight of the Pauli string to be implemented. 
In that case, the heuristic reverts to the first appearance of that state and then switches to an aggressively greedy approach where it selects an operation that is beneficial for implementing the Pauli string, even if this approach does not reduce the distance metric. 
After performing this greedy step, the heuristic switches back to reducing the distance metric. 
These heuristic choices ensure a solution is eventually attained, but aim to maximize reductions in Pauli weight in implementing the Pauli string.\\ 

\begin{algorithm}
\label{algo 2}
  \caption{Hardware Aware Heuristic} 
  \begin{algorithmic}[1]
  \REQUIRE length k Pauli word $\Omega_k$, index $ndx$, coupling list $C$, distance matrix $D$
  \STATE $reduction = []$
  \STATE $elapsed\_states = []$
  \STATE $switch = False$
  \WHILE{$Pauli\_weight(\Omega_k[ndx])\:!= 1$ }
    \STATE $operations\leftarrow gen\_ops(\Omega_k, ndx, D, C)$
    \STATE $ op \leftarrow best\_operation(\Omega_k, ndx, D,switch)$
    \STATE $\Omega_{new} = op\;\Omega_k\;op^{\dagger}$
    \IF{$\Omega_{new}$ not in $elapsed\_states$}
        \STATE $\Omega_{k}\leftarrow \Omega_{new}$
        \STATE $reduction.append(op)$
        \STATE $elapsed\_states.append(\Omega_k)$
    \ELSE
        \STATE $\Omega_k \leftarrow \Omega_{new}$
        \STATE $switch \leftarrow True$
        \STATE $index = elapsed\_states.index(\Omega_k)$
        \STATE $reduction \leftarrow reduction[:index]$
        \STATE $elapsed\_states \leftarrow elpsed\_states[:index]$
        \STATE $skip$
    \ENDIF
    \STATE $switch \leftarrow False$
  \ENDWHILE
  \STATE $\Omega_{k-m}, single\_q \leftarrow prune(\Omega_k)$
  \STATE $reduction.append(single\_q)$
  \RETURN $\Omega_{k-m}, reduction$
  \end{algorithmic}
\end{algorithm}
\vspace{1.85em}
\subsection{Additional MonteQ Information}
\noindent We discuss a few additional details concerning MonteQ that we find important for good performance.
MonteQ's circuit representations are based on the Qiskit framework~\cite{javadiabhari2024quantumcomputingqiskit}. 
A feature of particular interest is how $U_{CL}$, the product of tail Cliffords (see Subsection~\ref{circ op}), is synthesized. 
Because Clifford circuits can be optimized separately, MonteQ employs Qiskit's \textit{GreedySynthesisClifford} optimization pass to find an optimized form for $U_{CL}$.\\ 
\noindent In terms of MCTS, we note that an important practical consideration is how to handle a situation where the Pauli word is so large that no terminal node is ever reached by the exploration step within a budgeted number of iterations. 
To resolve this concern, our implementation of MCTS stores the solutions produced from the rollout policy at each simulation step, which reach a terminal node by the definition of rollout. 
Although reached greedily, these simulated solutions nonetheless define valid Pauli networks.
MonteQ  returns the solution, from among all terminal nodes regardless of whether they were reached by exploration or simulation, with the lowest CNOT count after $U_{CL}$ is optimized. 
This ensures that a solution is always provided, while also leveraging the MCTS framework's guidance for which leaf nodes to do rollout on. 
When multiple circuits in terminal nodes exhibit the same CNOT count, the total implemented circuit depth is used to break ties. \\

\section{Evaluation}
\label{sec:evaluation}
\subsection{Benchmarks and Implementation}

\noindent \textbf{Benchmarks}: We employ a set of benchmark unitaries, displayed in Table~\ref{tab:bench}. The molecular hamiltonians were acquired from PennyLane~\cite{Bergholm:2018cyq} from its Quantum Data dataset, specifically in the sto-3g basis set\cite{aszabo82:qchem}. The UCC ansatz were acquired using PennyLane to generate the necessary physics via its qml.qchem package and Openfermion~\cite{McClean:2017ims} to express this information as qubit operators.\\
\noindent The spin and fermion models are on square lattices of x rows and y columns. We acquired these dynamic models from Qiskit Nature~\cite{the_qiskit_nature_developers_and_contrib_2023_7828768}. We opted to convert the spin models to qubit operations using Qiskit Nature's LinearMapper and the fermion models were converted using the Jordan Wigner transformation~\cite{PhysRevA.98.022322}, also present within Qiskit Nature. Furthermore, we opted to include a wider range of qubit sizes for the Fermi-Hubbard and Heisenberg Models, in our logical regime tests, because of recent interest in the performance of dynamic system models on all-to-all connected systems~\cite{Srinivasan:2024fvq}\\
\textbf{Software Implementation}: MonteQ is implemented in Python 3.13 on top of Qiskit~\cite{javadiabhari2024quantumcomputingqiskit}. We implement it here \cite{github_code}.

\begin{table}[h!] 
\caption{Benchmarks}
    \begin{center}
        \begin{tabular}{ |c|*{2}{c|}}
            \hline
            Type & Name & Qubits \\
            \hline
            Spin Hamiltonian& Heisenberg Model (x,y) & 2xy\\
            \hline
            Fermion Hamiltonian& Fermi-Hubbard Model (x,y)& 2xy\\
            \hline
            Chemistry& LiH/H$_{2}$O/NH$_3$ &10/12/14\\
            \hline
            Variational Ansatz&UCC (\#electrons, \#orbitals)&\#orbitals\\
            \hline
            
        \end{tabular}
    \end{center}
 \label{tab:bench}
\end{table}

\subsection{Results}

\noindent \textbf{Single Iteration}: The results in this section focus on a single iteration of MCTS when using MonteQ. This is done to assess MonteQ's most basic output. We present the data in the Unitary Preserving logical regime in Table~\ref{tab:large tab}. 
The table presents the number of CNOTs in the circuits produced by MonteQ and the total time a single iteration takes in seconds. 
We  compare MonteQ against Qiskit's implementation of Rustiq and QuCLEAR~\cite{10946817} and present the percentage reductions MonteQ obtains for each benchmark. 
For maximal fairness, both comparators have the \textit{GreedySynthesisClifford} optimization pass applied to their tail Cliffords.
For QuCLEAR, we employ its Clifford Extraction algorithm and have the final Clifford optimized using the \textit{GreedySynthesisClifford} optimization pass instead of their Clifford Absorption algorithm.\\

\noindent\underline{\emph{CNOT Gate Count}}: In Table~\ref{tab:large tab}, MonteQ outperforms Qiskit-Rustiq in 17 of the 18 benchmarks and QuCLEAR in 13 of the 18 benchmarks. It showed a maximum possible improvement of 51.6\% (average 23.5\%) and 68.5\% (average 16.9\%) for Qiskit-Rustiq and QuCLEAR respectively.\par
\noindent This improvement makes it clear that our rollout policy, combined with the logical regime heuristic, is effective for circuit synthesis. The greatest losses MonteQ experiences are within the UCC ansatz. We believe this is because, in this case, MonteQ does not take advantage of MCTS and performs poorly on benchmarks with multiple complex Pauli strings. Here complex means they contain a larger number of nonidentity Paulis. For this reason, benchmarks containing sparser Pauli strings, like the dynamic models, show the greatest improvement for single iteration MonteQ.\par
\noindent\underline{\emph{Time}}: To assess the time taken to produce a solution, we included the time taken for the \textit{GreedySynthesisClifford} optimization pass. Table~\ref{tab:large tab} shows that single iteration MonteQ outperforms QuCLEAR in compilation time. This coupled with the improvements in CNOT gate count show that, without the use of MCTS, MonteQ is already a competitive compiler.\\

\begin{table*}[]
    \centering
    \caption{Single MCTS Iteration Comparisons}
    \begin{tabular}{ |c|*{8}{c|}}
        \hline
         \multirow{2}{*}{Name}&\multicolumn{5}{|c|}{\#CNOT}& \multicolumn{3}{|c|}{Time (sec)}\\
         \cline{2-9}
        &MonteQ&Qiskit-Rustiq&\%Reduction(Rustiq)&QuCLEAR&\%Reduction(QuCLEAR)&MonteQ&Qiskit-Rustiq&QuCLEAR\\
        \hline
        Fermi-Hub(2,3)&119&139&14.4\%&111&-7.21\%&0.321&0.0170&1.39\\
        \hline
        Fermi-Hub(3,3)&213&274&22.3\%&322&33.9\%&0.453&0.0240&4.12\\
        \hline
        Fermi-Hub(4,3)&338&489&30.9\%&547&38.2\%&0.639&0.0353&8.46\\
        \hline
        Fermi-Hub(4,5)&736&1076&31.8\%&1248&41\%&2.14&0.0832&37.2\\
        \hline
        Fermi-Hub(5,5)&933&1275&26.8\%&2529&63.1\%&3.53&0.118&85.4\\
        \hline
        Fermi-Hub(6,5)&1091&1704&36\%&3461&68.5\%&5.80&0.164&163\\
        \hline
        Heisenberg(2,2)&81&96&15.6\%&92&12\%&0.111&0.0197&2.20\\
        \hline
        Heisenberg(2,3)&154&176&12.5\%&158&2.53\%&0.232&0.0762&4.32\\
        \hline
        Heisenberg(3,3)&260&344&24.4\%&265&1.89\%&0.541&0.0387&11.8\\
        \hline
        Heisenberg(4,5)&654&986&33.7\%&829&21.1\%&3.54&0.165&87.5\\
        \hline
        Heisenberg(5,5)&809&1424&43.2\%&1038&22.1\%&6.48&0.268&159\\
        \hline
        Heisenberg(6,5)&967&1997&51.6\%&1297&25.4\%&14.5&0.408&257\\
        \hline
        LiH&561&668&16\%&539&-4.08\%&2.34&0.0753&5.56\\
        \hline
        H$_2$O&1089&1372&20.6\%&1130&3.63\%&9.44&0.231&14.7\\
        \hline
        NH$_3$&2553&3913&34.8\%&2926&12.7\%&64.7&3.27&46.3\\
        \hline
        UCC 2 4&26&27&3.7\%&25&-4\%&0.0328&0.00898&0.182\\
        \hline
        UCC 4 8&315&339&7.08\%&296&-6.42\%&0.919&0.0367&3.76\\
        \hline
        UCC 6 12&1965&1932&-1.71\%&1714&-14.6\%&25.7&0.550&29.6\\
        \hline
    \end{tabular}
    
    \label{tab:large tab}
\end{table*}

\noindent\textbf{Multiple Iterations}: In order to take advantage of MCTS, the number of iterations must be increased. As we stated in subsection ~\ref{circ op}, the number of ways to synthesize a circuit is factorial in the number of Pauli strings($K$). This implies that the tree size is also factorial in $K$. To get the optimal answer for our method it would require traversing the whole tree but that is intractable. For example, UCC\_2\_4, our smallest problem, has 12 Pauli strings. However, we can show that even for lower iteration numbers, improvement is possible.\par
\noindent\underline{\emph{CNOT Gate Count}}: We present the CNOT gate count results of MonteQ, in the Unitary Preserving logical regime, after 200 iterations in Table~\ref{tab: multiple iteration}. The percentage reductions presented are against MonteQ's single iteration results:

\begin{table}[ht] 
\caption{Multiple MCTS Iterations}
    \begin{center}
        \begin{tabular}{ |c|*{2}{c|}}
            \hline
            Name & 200 Iterations & \%Reduction against single iteration\\
            \hline
            Fermi-Hub(2,3)&104&12.6\%\\
            \hline
            Fermi-Hub(3,3)&201&5.63\%\\
            \hline
            Fermi-Hub(4,3)&299&11.5\%\\
            \hline
            Fermi-Hub(4,5)&595&19.2\%\\
            \hline
            Fermi-Hub(5,5)&776&16.8\%\\
            \hline
            Fermi-Hub(6,5)&1045&4.22\%\\
            \hline
            Heisenberg(2,2)&75&7.40\%\\
            \hline
            Heisenberg(2,3)&136&11.7\%\\
            \hline
            Heisenberg(3,3)&220&15.4\%\\
            \hline
            Heisenberg(4,5)&596&8.87\%\\
            \hline
            Heisenberg(5,5)&753&6.92\%\\
            \hline
            Heisenberg(6,5)&939&2.9\%\\
            \hline
            LiH&517&7.84\%\\
            \hline
            H$_2$O&1039&4.59\%\\
            \hline
            NH$_3$&2553&0\%\\
            \hline
            UCC 2 4&18&30.8\%\\
            \hline
            UCC 4 8&289&8.25\%\\
            \hline
            UCC 6 12&1745&11.2\%\\
            \hline

        \end{tabular}
    \end{center}

 \label{tab: multiple iteration}
\end{table}

\noindent The results show that MonteQ averages an improvement of 10.3\% across all benchmarks with a maximum of 30.8\% . Furthermore, having 200 iterations allows MonteQ to outperform Qiskit-Rustiq completely. As expected, the largest changes occurred in relatively smaller systems which would have smaller trees. The lowest change occurred in NH$_3$ which is the largest overall system with 1389 Pauli strings. \par

\begin{figure}[htbp]
\centerline{\includegraphics[width=87mm]{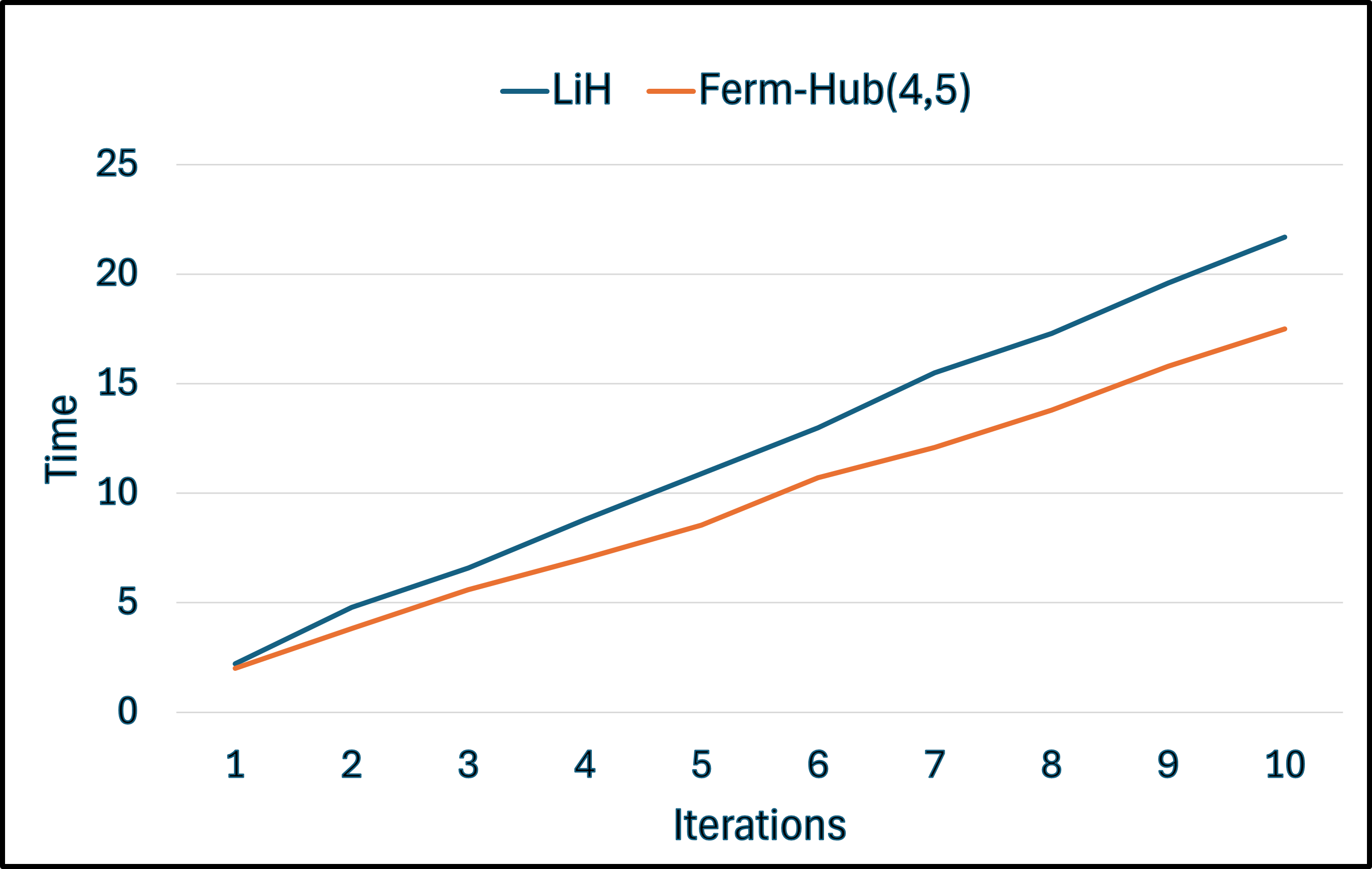}}
\caption{Relationship between Time taken and Iteration number}
\label{fig time vs iter}
\end{figure}

\noindent\underline{\emph{Time}}: To assess the relationship between the time MonteQ takes to acquire a solution and the number of iterations, we present Fig~\ref{fig time vs iter}. To create this figure, we used the Hamiltonians for LiH and the Fermi-Hubbard(4,5) Model and ran them each for up to 10 iterations in the Unitary Preserving logical regime. These two Hamiltonians were used because they have similar single iteration times. 

\begin{table*}[ht] 
\caption{Hardware-Aware CNOT Gate Count Results}
    \begin{center}
        \begin{tabular}{ |c|*{7}{c|}}
            \hline
            Name & MonteQ &Tetris&\% Reduction(Tetris)& TKet&\% Reduction(TKet)&Qiskit&\% Reduction(Qiskit)\\
            \hline
            Fermi-Hub(2,3)&190&188&-1.06\%&217&12.4\%&306&37.9\%\\
            \hline
            Fermi-Hub(3,3)&489&402&-2.16\%&582&16\%&556&12.1\%\\
            \hline
            Fermi-Hub(4,3)&939&680&-38.1\%&1164&19.3\%&1137&17.4\%\\
            \hline
            Heisenberg(2,2)&206&224&8.04\%&243&7.82\%&231&10.8\%\\
            \hline
            Heisenberg(2,3)&367&392&6.38\%&424&13.4\%&426&13.8\%\\
            \hline
            Heisenberg(3,3)&727&672&-8.18\%&946&23.2\%&807&9.91\%\\
            \hline
            LiH&1870&1930&3.11\%&3102&39.7\%&2888&35.2\%\\
            \hline
            H$_2$O&3964&4644&14.6\%&9118&56.5\%&7342&46\%\\
            \hline
            UCC 4 8&1037&1312&21\%&1437&27.8\%&1800&42.4\%\\
            \hline
            UCC 6 12&7162&9024&20.6\%&10562&32.2\%&12448&42.5\%\\
            \hline

        \end{tabular}
    \end{center}

 \label{tab: hardware}
\end{table*}

\begin{table}[ht] 
\caption{Unitary Modifying Results}
    \begin{center}
        \begin{tabular}{ |c|*{3}{c|}}
            \hline
            Name & MonteQ & Qiskit-Rustiq&\% Reduction \\
            \hline
            Fermi-Hub(2,3)&87&116& 25\%\\
            \hline
            Fermi-Hub(3,3)&190&255&25.5\%\\
            \hline
            Fermi-Hub(4,3)&289&352&17.9\%\\
            \hline
            Fermi-Hub(4,5)&584&686&14.9\%\\
            \hline
            Fermi-Hub(5,5)&828&944&12.3\%\\
            \hline
            Fermi-Hub(6,5)&1029&1134&9.26\%\\
            \hline
            Heisenberg(2,2)&74&103&28.2\%\\
            \hline
            Heisenberg(2,3)&132&193&31.6\%\\
            \hline
            Heisenberg(3,3)&228&407&44\%\\
            \hline
            Heisenberg(4,5)&622&1220&49\%\\
            \hline
            Heisenberg(5,5)&821&2064&60.2\%\\
            \hline
            Heisenberg(6,5)&993&2496&60.2\%\\
            \hline
            LiH&405&367&-10.4\%\\
            \hline
            H$_2$O&833&1135&26.6\%\\
            \hline
            NH$_3$&2160&3440&37.2\%\\
            \hline
            UCC 2 4&17&23&26\%\\
            \hline
            UCC 4 8&245&254&3.54\%\\
            \hline
            UCC 6 12&1476&2042&27.7\%\\
            \hline

        \end{tabular}
    \end{center}

 \label{tab: logic II}
\end{table}

\noindent From the figure, it is clear that the relationship between time taken and the number of iterations is roughly linear. This means that one can estimate the total time a certain number of iterations will take given the single iteration time. Thus, MonteQ can balance time taken against gains in CNOT gate count by setting the number of iterations.\\
\\
\noindent\textbf{Unitary Modifying}: Not preserving the unitary presents advantages because it allows more opportunities for Pauli strings to be co-reduced. In the Unitary Preserving regime, our heuristics focus on removing Pauli strings up to commutation. Without needing to care about commutation, the heuristics in MonteQ can remove any reduced Pauli strings in the problem.\par
\noindent To illustrate MonteQ's ability to disregard commutation of Pauli strings in a problem, we present the Unitary Modifying CNOT gate count results in Table~\ref{tab: logic II}. In this table, we compare MonteQ with Qiskit-Rustiq, using its own Unitary Modifying optimization. The comparison is performed in the logical regime and run for 200 iterations: 

\noindent The results show that MonteQ outperforms Qiskit-Rustiq in 17 out of the 18 benchmarks. It does so with a maximum  improvement of 60.2\% (27.2\% average). This shows that in both the Unitary preserving and modifying regimes, MonteQ is very competitive.\par
\noindent The results show that the Unitary Modifying regime improves on the Unitary Preserving regime for MonteQ as expected. With the freedom to remove any Pauli string in the problem, more co-reduced Paulis strings are removed in one go.\\  
\\
\noindent \textbf{Variation in Heuristic}: MonteQ has the ability to apply different heuristics for different needs. To this effect, we seek to assess its performance in CNOT Gate count in the hardware-aware regime.
\noindent In this situation, not all qubits are connected and so operations can not be applied indiscriminately. Some of the most prominent of these are IBM's superconducting qubit quantum computers\cite{AbuGhanem:2024atm}.\par
\noindent We compare the CNOT gate counts against Tetris~\cite{Jin:2023mil}, TKet ~\cite{Sivarajah:2020lfo} and Qiskit's standard transpiler set to an optimization level of 3. MonteQ is run, in the Unitary Preserving regime, for 200 iterations. MonteQ, Tetris, TKet and Qiskit all use the heavy-hex architecture of IBM's FakeManhattanV2. Specifically, MonteQ and Qiskit are tested using Qiskit's DenseLayout function to map the logical qubits to actual hardware qubits. Qiskit's transpiler at optimization level 3 endeavors to find better physical qubit to logical qubit mappings reduce the SWAP gates used. This means even though it starts with the DenseLayout, it may not complete in that layout. We present the results for 10 benchmarks in Table \ref{tab: hardware}.

\noindent The results show that MonteQ outperforms Tetris in 6 of the 10 benchmarks. It has a maximum improvement of 21\% (average 2.42\%). The most successful benchmarks are the molecular Hamiltonians and UCC ansatz. We believe this is because both of these cases have Pauli words with "denser", that is, containing more non-identity Paulis, Pauli strings. Within the hardware aware heuristic, if a Pauli string has many identity elements, SWAPs will have to be used to reduce the distance metric. This leads to larger CNOT count for benchmarks with "sparser" Pauli strings. It is for this reason, we believe that the sparser dynamic models perform poorly with our heuristic against Tetris. Against TKet, however, MonteQ outperforms in all the benchmarks. It has a maximum improvement of 56.5\% (average 24.8\%). The lowest improvements are found in the dynamic models which further confirms that the heuristic struggles against sparser Pauli strings. Lastly, MonteQ outperforms Qiskit's transpiler in all the benchmarks. It has a maximum improvement of 42.5\% (average 26.8\%).  The Qiskit comparison reinforces the idea that MonteQ improves on benchmarks with denser Pauli strings.  Overall, MonteQ is competitive because of its performance against Tetris, TKet and Qiskit. \\

\noindent\textbf{Future Work}: The overall results show that MonteQ is a competitive compiler. In this section we discuss future improvements to MonetQ that we have planned: 

\noindent\emph{Scalability} - As of now MonteQ struggles with time constraints, especially for larger systems. This is mainly because of the factorially large search space that exists. To combat this, truncation of the search space on the basis of Pauli weight is being considered. The main disadvantage to any truncation would be the reduction of possibly better optimization paths. Randomizing the truncation may improve this, but also runs the risk of producing nondeterminisitc results. A full solution would be to modify MonteQ to have different optimization tiers to include different levels of truncation.

\noindent\emph{Extension to General Circuits}: If some general circuit can be expressed as a sequence of Pauli rotations, it can be fed into MonteQ. This implies that MonteQ can be used on a class of general circuits outside of Quantum Simulation. A natural next step for MonteQ is adding the extension to convert and resynthesize general circuits. 

\section{Conclusion}
\label{sec:conclusion}
\noindent In this paper, we present MonteQ, a Monte Carlo Tree Search Based Quantum Circuit Synthesis Framework.  We show that a given synthesis problem can be expressed as a Markov Decision Process. From this, with the use of Monte Carlo Tree Search and a provided heuristic function, MonteQ can produce a solution circuit with the aim of reduced CNOT gate count. MonteQ has been shown to be versatile enough to tune heuristic choice, and amount of time taken while also balancing optimization of CNOT gate count. Our experimental results show that, given only simple heurisitic functions, MonteQ can outperform state-of-the-art compilers, especially within the logical regime. MonteQ also offers a flexible tradeoff between solution quality and runtime by adjusting the number of MCTS iterations, allowing the balance to be tuned based on user needs.

\section*{Acknowledgment}
\noindent This material is based upon work supported by the DOE-SC Office of Advanced Scientific Computing Research MACH-Q project under contract number DE-AC02-06CH11357. This research used resources of the Oak Ridge Leadership Computing Facility at the Oak Ridge National Laboratory, which is supported by the Advanced Scientific Computing Research programs in the Office of Science of the U.S. Department of Energy under Contract No. DE-AC05-00OR22725.

\bibliographystyle{IEEEtranS}
\bibliography{references}
\end{document}